\documentclass[rnote]{aa} % for the research notes
\usepackage{graphicx}
\usepackage{txfonts}
\begin{document}
   \title{The K-band properties of Seyfert 2 galaxies}

   \author{Zhixin Peng\inst{1}
          \and Qiusheng Gu\inst{1}
          \fnmsep\thanks{Visiting Scholar, Harvard-Smithsonian
          Center for Astrophysics, 60 Garden St., Cambridge, MA 02138, USA}
          \and Jorge Melnick\inst{2}
          \and Yinghe Zhao\inst{1}
          }

   \offprints{Q. Gu}

   \institute{Department of Astronomy, Nanjing University, Nanjing 210093, China \\
             \email{qsgu@nju.edu.cn;yhzhao@nju.edu.cn}
    \and European Southern Observatory, Alonso de Cordova 3107, Santiago, Chile \\
             \email{jmelnick@eso.org}
             }

   \date{Received ~~~~~~~~~~~~; accepted ~~~~~~~~~~~~~~~}

% \abstract{}{}{}{}{}
% 5 {} token are mandatory

  \abstract
   {}
   {It is well known that the [O {\sc iii}]$\lambda$5007 emission
 line and hard X-ray(2-10keV) luminosities are good indicators of
 AGN activities and that the near and mid-infrared emission of
 AGN originates from re-radiation of dusty clouds heated
 by the UV/optical radiation from the accretion disk. In this paper
 we present a study of the near-infrared K-band (2.2$\mu$m)
 properties for a sample of 65 Seyfert 2 galaxies.}
   {By using the AGN/Bulge/Disk decomposition technique, we analyzed
    the 2MASS K$_{\rm S}$-band images for Seyfert 2 galaxies in order
    to derive the K$_{\rm S}$-band magnitudes for the central engine,
    bulge, and disk components.}
   {We find that the K$_{\rm S}$-band magnitudes of the central
   AGN component in Seyfert 2 galaxies are tightly correlated with
   the [O {\sc iii}]$\lambda$5007 and the hard X-ray luminosities,
   which suggests that the AGN K-band emission is also an excellent
   indicator of the nuclear activities at least for Seyfert 2
   galaxies. We also confirm the good relation between the central
   black hole masses and bulge's K-band magnitudes for Seyfert 2s.}
   {}


   \maketitle
%
%________________________________________________________________

\section{Introduction}

 In the standard unified scheme, Seyfert 1 and 2 galaxies are
 intrinsically the same objects; the absence of broad Balmer
 emission lines in Seyfert 2s is due to the obscuration by a
 pc-scale dusty torus oriented along the line of sight (Krolik \&
 Begelman 1988; Antonucci 1993). The primary evidence for this model
 is the detection of polarized broad Balmer emission lines in dozens
 of Seyfert 2 galaxies, which have betrayed the hidden broad line
 regions via scattering off electrons and/or dust located above the
 hole of the torus (Antonucci \& Miller 1985; Tran 1995 and 2001;
 Young et al. 1996; Heisler et al. 1997; and Moran et al. 2000).

 One natural prediction of the unified model is that such a dusty
 torus surrounding the central engine will absorb the UV/optical
 radiation from the accretion disk and re-radiate it in the infrared,
 as the sublimation temperature of graphite grains is in the range
 1500 $-$ 1800 K. According to the Wien law, $T\lambda_{peak}(\mu
 m) = 2898$, the peak wavelength of re-radiation from the dusty
 torus should be around 2$\mu$m. In fact, it is well known that
 Seyfert galaxies are strong near and mid-infrared sources (Rieke
 1978; Neugebauer et al. 1979). Pier \& Krolik (1992) performed the
 first calculation of thermally re-radiated infrared spectra of the
 compact dust torus, and found good agreement between the model
 predictions and observations in Pier \& Krolik (1993), which
 suggested that near infrared emission in Seyfert galaxies mainly
 arise from the re-radiation of dusty torus heated by UV/optical
 light from the central engine. Recently, Suganuma et al. (2005)
 presented reverberation measurements of several well-known Seyfert
 1 galaxies by monitoring flux variations in the optical (UBV) and
 near-infrared (JHK) bands. They find a clear time-delayed behavior
 of the K-band flux variations relative to the V-band. From the flux
 variation gradients they derive (H $-$ K) color temperatures of
 1500 $-$ 1800 K, and suggest that the bulk of the K-band emission
 originates from thermal emission of dust grains in the optically
 thick torus (see also Glass 2004). Thus the K-band emission should
 also be a good indicator of AGN activity.

 The recent, comprehensive Two Micron All Sky Survey (2MASS) has
 uniformly scanned the entire sky in three near-infrared (NIR) (J
 1.25$\mu$m, H 1.65$\mu$m, and K$_{\rm S}$ 2.17$\mu$m) bands.
 % with the photometric measurement accuracy of 5\% for point sources and 10\% for extended objects.
 The point-source sensitivity limits (10$\sigma$) are J=15.8 (0.8
 mJy), H=15.1 (1.0 mJy), and K$_{S}$=14.3 (1.3 mJy) mag. The extended
 source sensitivity (10$\sigma$) is $\sim$ 1 mag brighter than the
 point source
 limits\footnote{http://www.ipac.caltech.edu/2mass/releases/spr99/doc/test/jarrett2/intro.html}.
 The 2MASS public catalog contains $>$ 1,000,000 galaxies and thus
 provides the most comprehensive database for the study of NIR
 galaxy properties.

 In this paper we have performed  AGN/Bulge/Disk decomposition for a
 sample of Seyfert 2 galaxies by using 2MASS K$_{\rm S}$-band images
 in order to derive the near-infrared magnitudes of the central
 AGNs. In Sect. 2, we present a short description for our sample
 and present the results of decomposition in Sect. 3. We discuss
 our results in Sect. 4 and present conclusions in Sect. 5. By
 using the WMAP data, Spergel et al. (2003) have well determined the
 Hubble constant to be H$_0$= 71 km s$^{-1}$ Mpc$^{-1}$, which will be
 used  throughout this paper.

\section{Data}

 We recently performed a systematic study of a large and homogeneous
 sample of 65 nearby Seyfert 2 galaxies observable from the southern
 hemisphere. A full description of the sample selection,
 observation, and data reduction is presented in Joguet et al.
 (2001). The star formation history and stellar populations in the
 central $\sim$ 200pc for these Seyfert 2s are derived by means of
 stellar population synthesis modelling (Cid Fernandes et al. 2004),
 and Gu et al. (2005) analyzed the "pure" emission line spectra
 obtained by subtracting the synthetic stellar contribution to
 derive pure emission-line spectra. Here we use the 2MASS K$_{\rm
 S}$-band images of all Seyfert 2 galaxies in our sample\footnote{http://irsa.ipac.caltech.edu/applications/2MASS/PubGalPS/},
 to present a study of NIR properties for these galaxies. The plate
 scale of the 2MASS K$_{\rm S}$-band image is 1.0 arc-second per
 pixel.

\begin{table*}[h]
\caption{Properties of Seyfert 2 galaxies}
\begin{tabular}{lccccccc}
\hline\noalign{\smallskip} Galaxy  &  D   &   $\sigma$ &  m$_{\rm
K,AGN}$& m$_{\rm K,bulge}$  & m$_{\rm K,disk}$
& log L$_{\rm [OIII]}$  &   log L$_{\rm 2-10keV}$\\
 & (Mpc)& (km s$^{-1}$) & (mag) & (mag) & (mag) & (ergs s$^{-1}$) & (ergs s$^{-1}$) \\
\hline\noalign{\smallskip}
   ESO 103-G35   &   56.1   &   114.0   &   12.08   &   11.10   &   12.29   &   40.65   &   42.84\\
   ESO 104-G11   &   63.9   &   130.0   &   11.98   &   11.51   &   10.99   &   40.17   &        \\
   ESO 137-G34   &   38.7   &   133.4   &    9.63   &    9.20   &    8.11   &   40.80   &        \\
   ESO 138-G01   &   38.6   &    80.0   &   11.70   &   10.71   &   10.53   &   40.90   &   $>$41.51$^a$\\
   ESO 269-G12   &   69.7   &   160.7   &   13.19   &   11.44   &   10.69   &   39.53   &        \\
   ESO 323-G32   &   67.5   &   130.6   &   11.84   &   10.31   &   12.60   &   40.77   &        \\
   ESO 362-G08   &   67.4   &   154.2   &   11.78   &   10.07   &    7.14   &   40.19   &        \\
   ESO 373-G29   &   39.5   &    92.3   &   12.64   &   10.69   &   10.41   &   40.21   &        \\
   ESO 381-G08   &   46.2   &    99.7   &   13.67   &   10.24   &   11.30   &   41.00   &        \\
   ESO 383-G18   &   52.4   &    92.3   &   12.90   &   11.99   &    7.39   &   40.35   &        \\
\\
   ESO 428-G14   &   22.9   &   119.7   &   11.28   &    9.14   &   13.07   &   40.52   &   $>$40.69$^a$\\
   ESO 434-G40   &   35.0   &   144.9   &   10.88   &   10.18   &    9.92   &   40.28   &   43.11\\
   Fair 334   &   78.3   &   104.2   &   13.08   &   11.88   &   10.90   &   39.93   &   43.00\\
   Fair 341   &   67.9   &   121.9   &   13.06   &   10.74   &   10.93   &   40.79   &   43.08\\
   IC 1657   &   50.5   &   143.2   &   12.71   &   10.30   &   10.24   &   38.90   &        \\
   IC 2560   &   41.2   &   143.9   &   12.22   &    9.17   &   10.09   &   40.46   &   $>$40.99$^a$\\
   IC 5063   &   48.0   &   182.0   &   12.40   &    9.41   &    9.25   &   40.96   &   42.92\\
   IC 5135   &   68.2   &   143.2   &   12.17   &   10.18   &   10.17   &   41.15   &   $>$ 41.45$^a$\\
   IRAS F11215-2806   &   56.9   &    97.5   &   14.24   &   11.93   &   12.97   &   40.81   &        \\
   MCG +01-27-20   &   49.3   &    93.9   &   13.84   &   11.77   &   11.53   &   40.21   &        \\
\\
   MCG -03-34-64   &   69.8   &   155.2   &   11.24   &   11.13   &   10.26   &   41.86   &   42.58\\
   Mrk 897   &   111.2   &  133.0   &   12.23   &   13.34   &       0   &   40.13   &        \\
   Mrk 1210   &   56.9   &   114.0   &   12.69   &   11.60   &   10.55   &   41.05  &   $>$42.01$^a$\\
   Mrk 1370   &  103.8   &    86.1   &   13.73   &   13.18   &       0   &   40.45   &        \\
   NGC 424   &   49.2   &   142.6   &   10.34   &   11.23   &   10.04   &   41.78   &   $>$ 41.55$^a$\\
   NGC 788   &   57.5   &   162.9   &   11.93   &    9.77   &       0   &   40.16   &        \\
   NGC 1068   &   16.1   &   143.5   &    8.02   &    7.32   &    6.80   &   42.00   &   $>$41.03$^a$\\
   NGC 1125   &   46.2   &   104.7   &   12.05   &   11.38   &   10.77   &   39.84   &        \\
   NGC 1667   &   64.0   &   148.9   &   12.20   &    9.63   &   10.42   &   39.65   &   $>$42.39$^a$\\
   NGC 1672   &   19.0   &    96.8   &   11.54   &    8.95   &    7.82   &   38.88   &   41.21\\
\\
   NGC 2110   &   32.9   &   241.5   &   12.15   &    8.53   &   10.02   &   39.99   &   42.62\\
   NGC 2979   &   38.3   &   112.2   &   12.62   &   10.39   &   10.12   &   39.29   &         \\
   NGC 2992   &   32.5   &   171.8   &   11.33   &    8.96   &    8.47   &   40.46   &   41.76\\
   NGC 3035   &   61.4   &   161.4   &   12.06   &   12.41   &   10.03   &   40.41   &        \\
   NGC 3081   &   33.6   &   133.4   &   12.28   &   10.79   &    9.63   &   40.55   &   41.96\\
   \hline\noalign{\smallskip}
\end{tabular}
\end{table*}

\setcounter{table}{0}
\begin{table*}[h]
\caption{{\it continued}}
\begin{tabular}{lccccccc}
\hline\noalign{\smallskip} Galaxy  &  D   &   $\sigma$ & m$_{\rm
K,AGN}$& m$_{\rm K,bulge}$  & m$_{\rm K,disk}$
& log L$_{\rm [OIII]}$  &   log L$_{\rm 2-10keV}$\\
 & (Mpc)& (km s$^{-1}$) & (mag) & (mag) & (mag) & (ergs s$^{-1}$) & (ergs s$^{-1}$) \\
\hline\noalign{\smallskip}
   NGC 3281   &   45.1   &   160.0   &   13.24   &   10.22   &    8.54   &   40.26   &   42.84\\
   NGC 3362   &   116.7  &   103.5   &   13.68   &   11.05   &   13.19   &   41.59   &        \\
   NGC 3393   &   52.8   &   156.7   &   11.47   &   10.60   &    9.83   &   41.05  &   $>$41.13$^a$\\
   NGC 3660   &   51.8   &    95.5   &   12.39   &   10.51   &   11.57   &   40.63   &        \\
   NGC 4388   &   35.5   &   110.7   &   11.26   &    9.23   &    9.22   &   40.98   &   42.81\\
   NGC 4507   &   49.9   &   144.5   &   11.44   &    9.51   &   10.70   &   41.22   &   43.33\\
   NGC 4903   &   69.5   &   200.4   &   13.33   &   10.46   &   12.41   &   41.02   &        \\
   NGC 4939   &   43.8   &   154.9   &   12.18   &   11.50   &    9.04   &   40.75   &   42.01\\
   NGC 4941   &   15.6   &    98.2   &   11.25   &    9.01   &    9.22   &   39.99   &    40.95\\
   NGC 4968   &   41.6   &   121.1   &   12.05   &   10.75   &    9.44   &   40.49   &   $>$40.92$^a$\\
\\
   NGC 5135   &   57.9   &   142.6   &   11.11   &   10.17   &    9.98   &   40.57   &   $>$40.91$^a$\\
   NGC 5252   &   97.4   &   208.9   &   12.76   &   10.49   &   13.44   &   40.49   &   43.12\\
   NGC 5427   &   36.9   &   100.2   &   13.16   &    9.88   &    9.29   &   39.90   &        \\
   NGC 5506   &   26.1   &    97.9   &    9.63   &   10.74   &    9.02   &   40.74   &   42.94\\
   NGC 5643   &   16.8   &    92.9   &   11.17   &    7.65   &   11.25   &   39.91   &   $>$ 40.65$^a$\\
   NGC 5674   &  105.2   &   128.5   &   11.99   &   12.27   &   10.21   &   39.92   &   43.23\\
   NGC 5728   &   39.3   &   155.2   &   12.43   &    9.89   &    8.41   &   40.82   &        \\
   NGC 5953   &   27.7   &    92.9   &   12.88   &   10.03   &   10.38   &   39.31   &        \\
   NGC 6221   &   20.8   &   111.4   &   10.72   &    8.21   &    7.96   &   39.33   &        \\
   NGC 6300   &   15.6   &    99.8   &   11.22   &    9.65   &    6.15   &   39.44   &   41.28\\
\\
   NGC 6890   &   34.1   &   108.9   &   12.40   &    9.35   &   13.41   &   40.64   &         \\
   NGC 7172   &   36.7   &   190.0   &   12.18   &   10.23   &    8.89   &   39.00   &  \\
   NGC 7212   &   112.5  &   167.9   &   12.47   &   11.70   &    9.87   &   42.01   &   42.29\\
   NGC 7314   &   20.0   &    59.9   &   12.86   &   11.43   &    7.71   &   39.17   &   42.29\\
   NGC 7496   &   23.2   &   100.7   &   12.15   &    9.91   &    8.87   &   39.38   &   41.71\\
   NGC 7582   &   22.2   &   132.1   &   10.13   &    8.98   &    7.43   &   40.39   &   42.21\\
   NGC 7590   &   22.5   &    98.6   &   11.53   &    9.63   &    8.44   &   39.26   &   40.86\\
   NGC 7679   &   72.4   &    95.9   &   13.63   &   10.59   &   10.27   &   40.31   &   42.52\\
   NGC 7682   &   72.4   &   152.4   &   14.65   &   11.21   &    9.69   &   40.74   &   42.91\\
   NGC 7743   &   24.1   &    95.3   &   11.60   &    9.98   &    7.97   &   38.74   &   39.56\\
 \hline\noalign{\smallskip}
\end{tabular}

\noindent $^a$ Compton-thick source with column density $\ge$
10$^{24}$ cm$^{-2}$.
\end{table*}

\subsection{Light decompositions}

We applied the 2-dimensional bulge-disk decomposition program,
GALFIT {\sc version 2.0.3c} (Peng et al. 2002), to the 2MASS K$_{\rm
S}$-band images of these 65 Seyfert 2 galaxies. Specifically, in
order to separate the AGN component, we fitted each galaxy with a
combination of a point spread function (PSF) centered in the image,
an inclined exponential disk, and a bulge with a surface brightness
profile of $exp[-(r/r_s)^{1/n}]$ (Sersic 1968) with the index n
constrained to the range of 0.5 $\le$ n $\le$ 5.0.

 Ground-based images usually show near-Gaussian or Moffat-like
 PSFs (Peng et al. 2002). We generated an analytic
 profile of PSF for the 2MASS K$_{\rm S}$-band images using the
 standard tasks in IRAF\footnote{IRAF is distributed by the
 National Optical Astronomy Observatory, which is operated by the
 Association of Universities for Research in Astronomy, Inc.,
 under cooperative agreement with the National Science
 Foundation.}. Following Jarrett et al. (2000), we used a
 generalized, radially symmetric exponential function to describe
 the 2MASS K$_{\rm S}$-band PSF, which is:

\begin{equation}
f(r) = f_0  \rm{exp}
\left[-\left(\frac{r}{\alpha}\right)^{1/\beta}\right]
\end{equation}

 \noindent where $f_0$ is the central surface brightness, and
$\alpha$ and $\beta$ are free parameters. For the case of the 2MASS
survey, we adopted $\beta$ = 0.5 and $\alpha = \sqrt{\displaystyle
2}\sigma = 1.5''$, where $\rm \sigma=FWHM/2.354$, FWHM is the seeing
of the 2MASS images, and FWHM= $2.5''$ for the K$_{\rm S}$ images
{\footnote{http://spider.ipac.caltech.edu
/staff/roc/2mass/seeing/seesum.html}.

%--------------------------------------------------%figure 1
\begin{figure*}
   \centering
   \includegraphics[width=14cm]{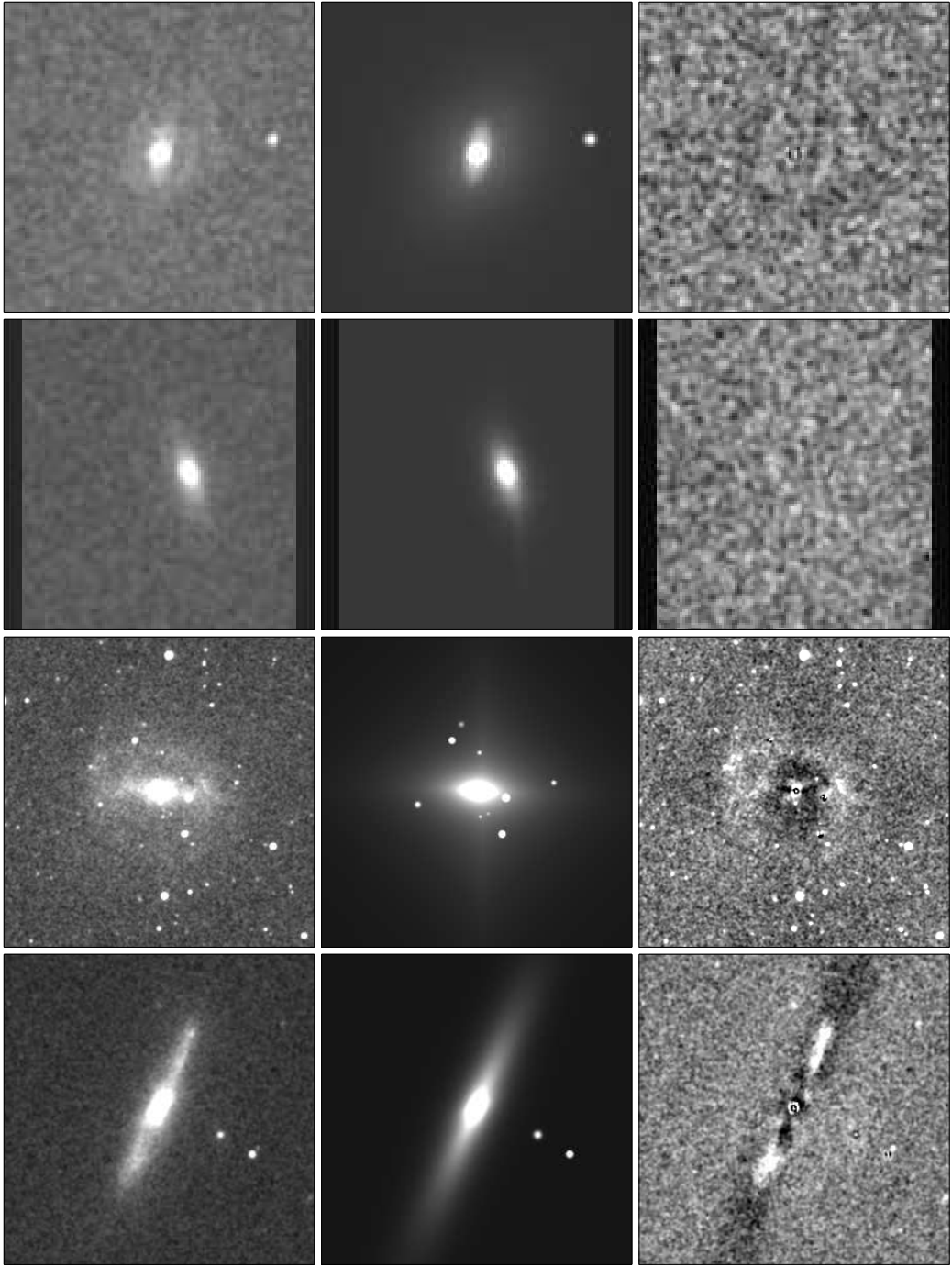}
\caption{Results of decomposition. As an example, we only show 4
representative Seyfert 2s with the smallest to the largest
$\chi^2$ (NGC 3035, NGC 5252, NGC 5643 and NGC 7582,from {\it top}
to {\it bottom}). For each source, we show the original image
({\it left}), the model ({\it middle}) fitted by GALFIT, and the
residual image ({\it right}).}
\end{figure*}
%--------------------------------------------------%figure 1

 Examples of the fitting results are shown in Fig.1 for 4
 representative Seyfert 2s covering the full range of $\chi^2$ (NGC
 3035, NGC 5252, NGC 5643, and NGC 7582 with $\chi^2$ of 0.012,
 0.013, 0.032, and  0.049, respectively). The standard measurement
 errors for AGN, bulge, and disk K$_{\rm S}$-band magnitudes are
 0.2, 0.3, and 0.5 mag, respectively. For NGC 5643 there appears to
 be a polar ring left after subtracting the bulge/disk components.
 This illustrates the high quality of the Bulge/Disk subtraction and
 may indicate that the ring reflects a dynamical interaction with
 another galaxy. The currently favored model for the formation of
 polar rings is suggested to be galactic mergers (Sparke \& Cox 2000;
 Bournaud \& Combes 2003; and Maccio, Moore \& Stadel 2006). For NGC
 7582, the residual flux after B/D subtraction is peculiar, which may
 indicate a warp or some other faint asymmetry in the disk. It is
 interesting to note that for three objects (Mrk 897, Mrk1370, and
 NGC 788 with morphological types of Scd, Sa, and SA(s)0/a,
 respectively),  there seems to be no need for exponential disk
 components to fit the K$_{\rm S}$-band images.

 The decomposed K$_{\rm S}$-band magnitudes of AGN, bulge, and disk
 are listed in Table 1, where for completeness we also tabulate
 the stellar velocity dispersions ($\sigma$) from
 Cid Fernandes et al. (2004), and the [O {\sc
 iii}]$\lambda$5007 emission line and the hard X-ray(2-10 keV)
 continuum luminosities from Gu et al. (2005).

\section{Results}

%\subsection{K$_{\rm S}$-band emission: a good indicator of AGN activities}

 In the past three decades many attempts were made to establish
 which are the truly isotropic emissions from Seyfert galaxies.
 Among various emissions at different wavelengths, the [O {\sc
 iii}]$\lambda$5007 and hard X-ray (2-10kev) continuum luminosities
 have been found to show similar distributions for Seyfert 1 and 2
 galaxies (Dahari \& De Robertis 1988; Keel et al. 1994; Mulchaey et
 al.  1994; Alonso-Herrero et al.  1997; Kauffmann et al. 2003),
 indicating isotropic emission and thus making them reliable tracers
 of the intrinsic nuclear activity. Alonso-Herrero, Ward \&
 Kotilainen (1997) compared the hard X-ray, the [O {\sc
 iii}]$\lambda$5007, and the small-aperture L-band (3.5$\mu$m)
 properties of low-redshift PG quasars, Seyfert 1, and 2 galaxies,
 and they found good correlations between the [O {\sc
 iii}]$\lambda$5007 and L-band properties and between the hard X-ray
 and L-band properties for Seyfert 1s and PG quasars.

%______________________________________________ Figure 2
   \begin{figure}
   \centering
   \includegraphics[width=8cm]{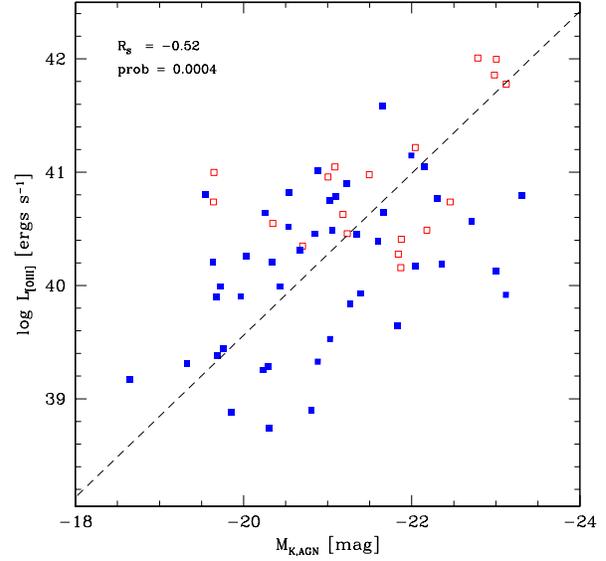}
   \caption{The AGN K$_{\rm S}$-band absolute magnitude versus [O
{\sc iii}]$\lambda$5007 luminosity. Red open-square symbols
correspond to Seyfert 2s with direct evidence of hidden broad-line
regions (HBLR) from spectropolarimetric observations. The dashed
line shows a ordinary least square (OLS) bisector fit to the data of
slope $-0.715\pm0.051$. The Spearman rank-order correlation
coefficient ($R_S$) is -0.52 and the null probability is 0.0004.}
              \label{Fig2}%
    \end{figure}
%______________________________________________ Figure 2

%______________________________________________ Figure 3
   \begin{figure}
   \centering
   \includegraphics[width=8cm]{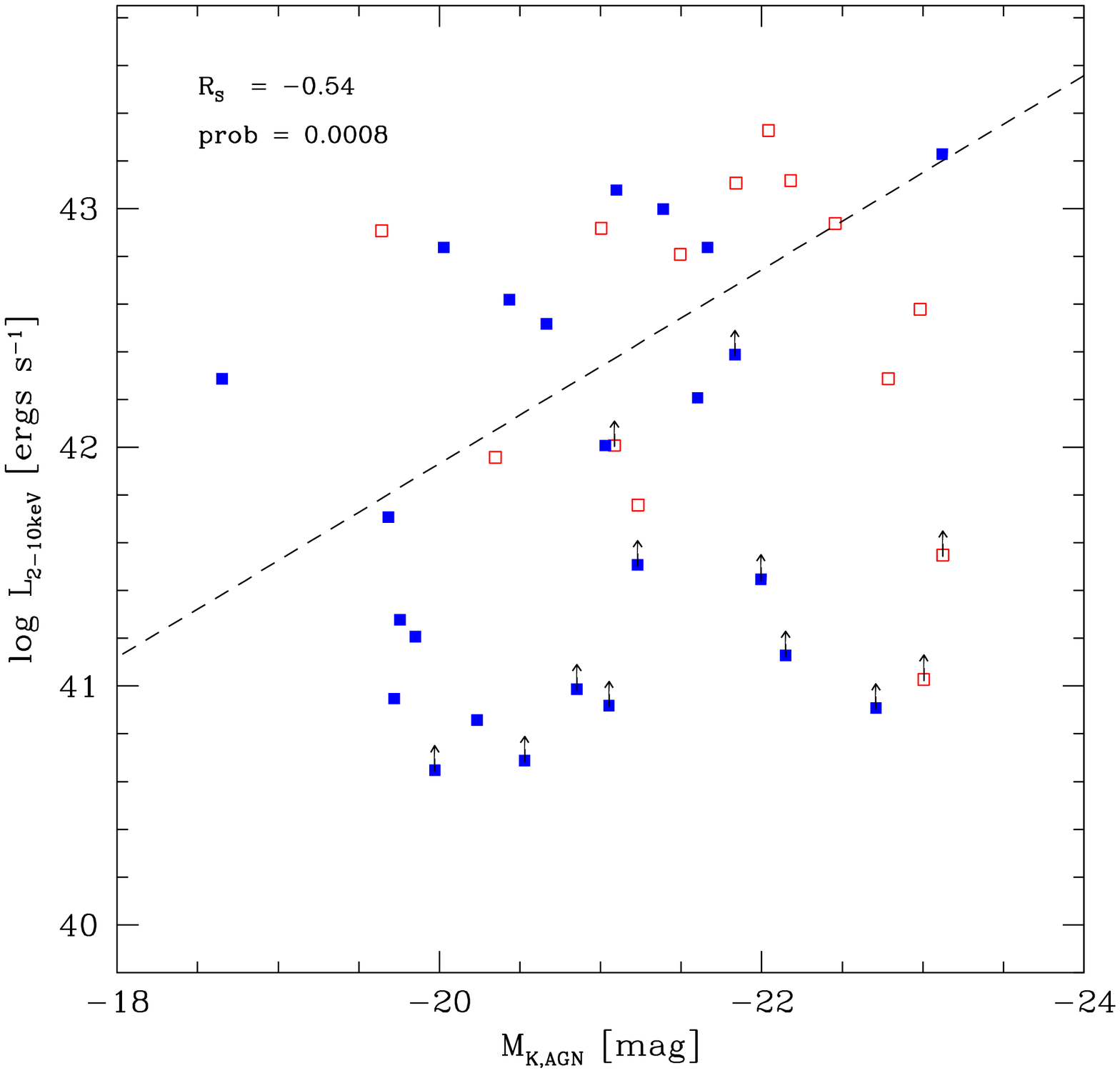}
   \caption{The AGN K$_{\rm S}$-band absolute magnitude versus 2-10 keV
hard X-ray luminosity. The dashed line shows an OLS bisector fit to
the data of slope $-0.406\pm0.124$. The Spearman rank-order
correlation coefficient ($R_S$) is -0.54 and the null probability is
0.0008. Symbols have the same meaning as in Fig. 2.}
              \label{Fig3}%
    \end{figure}
%______________________________________________ Figure 3

 Figure 2 shows the correlation between the K$_{\rm S}$-band
 absolute magnitudes of our Seyfert 2s with [O {\sc
 iii}]$\lambda$5007 luminosity. The red open squares represent
 objects where the spectropolarimetric observations detected the
 polarized broad emission lines (see also Gu et al. 2005), mainly
 from the compilation by Gu \& Huang (2002) plus new observations by
 Lumsden, Alexander \& Hough (2004). As both the AGN K$_{\rm
 S}$-band absolute magnitude ($\rm M_{\rm K,AGN}$) and [O {\sc
 iii}]$\lambda$5007 luminosity ($\rm L_{[OIII]}$) are well
 determined, in order to get the best fit between $\rm M_{\rm
 K,AGN}$ and $\rm L_{[OIII]}$, we used both the ordinary
 least-square (OLS) bisector method (Isobe et al. 1990) and the
 Buckley-James regression method from the Astronomy Survival
 Analysis package (ASURV v1.2; Isobe et al. 1985; 1986), and derived
 the same fitting results, which is $\rm \log L_{[OIII]} =
 (-0.715\pm0.051) \ M_{\rm K,AGN} + (25.260\pm1.092)$. Figure 2 also
 quotes the Spearman rank-order correlation coefficient ($R_{\rm
 S}$, Press et al. 1992) as -0.52, and a probability of $P_{null}
 = 0.0004$ for the null hypothesis of no correlation between $\rm
 M_{\rm K,AGN}$ and $\rm L_{\rm [OIII]}$.

 Similarly, Fig. 3 shows the correlation between the AGN K$_{\rm
 S}$-band absolute magnitude versus the hard X-ray 2-10keV
 luminosity. Twelve of our Seyferts  are Compton-thick sources with
 column densities larger than 10$^{24} cm^{-2}$, which are derived
 by fitting a simple power law with cold absorption to the X-ray
 spectrum. For Compton-thick objects, the direct hard X-ray continua
 are completely absorbed, so we cannot estimate their intrinsic hard
 X-ray luminosities. For these objects, the hard X-ray 2-10keV
 luminosities will be taken as lower limits. We have to derive the
 linear regression using the Buckley-James regression method from
 ASURV, which is $\rm \log L_{2-10keV} = (-0.406\pm0.124) \ M_{\rm
 K,AGN} + (33.762\pm2.594)$, with a Spearman rank-order correlation
 coefficient ($R_{\rm S}$) of $-0.54$ and a null probability of
 $P_{null} = 0.0008$.  As a worst case, if ALL lower limits were
 detected at their exact lower limit values, the correlation still
 holds but becomes much steeper. By using the same survival
 techniques, we derive the best fit to be $\rm \log L_{2-10keV} =
 (-0.912\pm0.056) \ M_{\rm K,AGN} + (22.647\pm1.237)$, with the
 Spearman rank-order correlation coefficient ($R_{\rm S}$) of
 $-0.30$ and a null probability of $P_{null} = 0.061$.

 Both Fig.2 and Fig.3 indicate that the AGN K-band magnitudes are
 tightly correlated with [O {\sc iii}]$\lambda$5007 and the hard
 X-ray(2-10keV) luminosities, both of which are considered to be good
 indicators of nuclear activities. Thus it suggests that the AGN
 K-band magnitude is also a good indicator of AGN activities for
 Seyfert 2 galaxies. We find in Fig.2 that Seyfert 2s with hidden
 broad-line regions clearly have much higher [O {\sc
 iii}]$\lambda$5007 luminosities and higher K$_{\rm S}$-band
 absolute magnitudes, indicating much powerful AGN activities
 as found by Gu \& Huang (2002).

 %which is $\rm log M_{BH} M_\odot = (1.21\pm0.13) \log L_{\rm K,Bulge} + (5.109\pm0.10)$

\section{Discussion}

 Since the seeing of 2MASS K$_{\rm S}$-band images is rather poor
 ($\sim 2.5''$), it is possible that part of the emission from AGN
 is mixed with that of the bulge. In order to check whether GALFIT did
 work in this situation, we used the Hubble Space Telescope NICMOS
 K-band images, which have much better spatial resolution than
 2MASS. Fortunately, 3 sources in our sample have NICMOS K-band
 data: NGC 2110, NGC 2992, and NGC 5643. The AGN magnitudes from
 NICMOS GALFIT fitting are 12.17, 12.03, and 10.96 mag,
 respectively. Our values from the  2MASS data are 12.15,
 11.33, and 10.69 mag, respectively. For NGC 2110 and NGC 5643,
 the fitting results are consistent within the typical fitting
 errors of GALFIT (see Sect. 2) with the NICMOS magnitudes.
 But for NGC 2992, which is an unabsorbed Seyfert 2 galaxy with a
 polarized broad H$\alpha$ emission (Panessa \& Bassani 2002; Rix et
 al. 1990), the magnitude difference is as large as 0.7 mag, which could
 be evidence of variability or perhaps of an aperture/PSF effect.

\subsection{Relation between BH mass and bulge luminosity}

 It is well known that there is a tight correlation between black
 hole masses and the bulge properties in spiral galaxies (Magorrian et
 al. 1998; Gebhardt et al. 2000; Ferrarese \& Merritt 2000). Marconi
 \& Hunt (2003) find that NIR bulge luminosities and BH masses are
 tightly correlated. For those galaxies with accurate BH masses, the
 spread of such a relation is nearly the same as the famous M$_{\rm
 BH}$ - $\sigma$ relation.

 Figure 4 shows the relation of bulge K-band absolute magnitudes and
 BH masses for our sample of Seyfert 2s. The BH mass is derived from
 the well-known M$_{\rm BH}$ - $\sigma$ relation, which is $\rm \log
 (M_{BH}/M_\odot) = (8.13 \pm 0.06) + (4.02\pm 0.32) \times
 \log(\sigma/200km s^{-1}) $ (Tremaine et al. 2002), where $\sigma$
 is the stellar velocity dispersion derived by Cid
 Fernandes et al. (2004). The dashed line is the best fit to the
 data by means of the OLS bisector method,
 which is $\rm log M_{BH} (M_\odot) = (1.199\pm0.080) \log (L_{\rm
 K,Bulge}/L_{\rm K,\odot}) - (5.184\pm0.831)$. It is very
 interesting to note that the relation between bulge K-band
 magnitude and BH mass for Seyfert 2s is nearly the same as that
 derived for normal galaxies with secure BH mass measurement (see
 Table 2 in Marconi \& Hunt 2003), which shows that host galaxies
 of Seyfert 2s obey the same relation as normal galaxies. However,
 our scatter is much larger than that of Marconi \& Hunt (2003), the
 main reason being that BH masses are well-measured in Marconi \& Hunt
 (2003), while we estimate BH masses via  the M$_{\rm BH}$ -
 $\sigma$ relation, and the stellar velocity dispersion was derived
 by stellar synthesis and thus has larger uncertainties.

%______________________________________________ Figure 4
   \begin{figure}
   \centering
   \includegraphics[width=8cm]{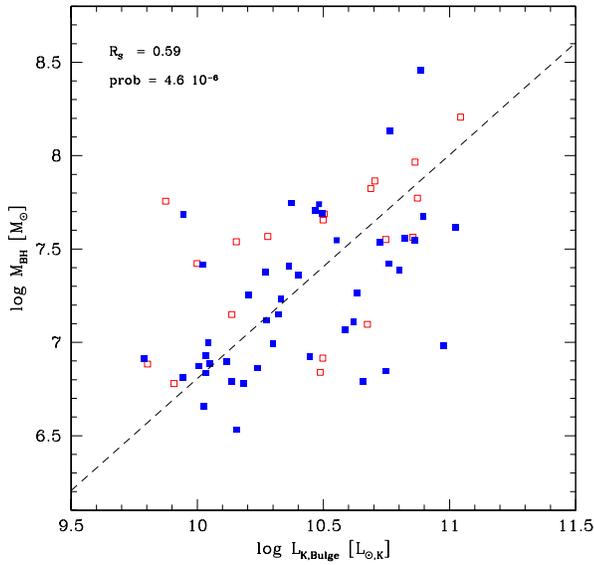}
   \caption{The bulge's K$_{\rm S}$-band absolute magnitude versus
the black hole mass. The dashed line shows an OLS bisector fit to
the data of slope $1.199\pm0.080$. The Spearman rank-order
correlation coefficient ($R_S$) is 0.59 and the null probability 4.6
10$^{-6}$. Symbols have the same meaning as in Fig. 2.}
              \label{Fig4}%
    \end{figure}
%______________________________________________ Figure 4

\section{Conclusion}
In this paper, we present a study of NIR properties for a sample of
Seyfert 2 galaxies. After decomposing the 2MASS K$_{\rm S}$-band
image in the AGN, bulge, and an exponential disk, we derived the
K$_{\rm S}$-band magnitudes for each component. We find that the AGN
K$_{\rm S}$-band magnitudes are tightly correlated with the
luminosities of [O {\sc iii}]$\lambda$5007 and the hard X-ray
emission, which suggests that the K-band emission is also an
excellent indicator of nuclear activity, at least for Seyfert 2
galaxies. We also confirm the good relation between the central
black hole masses and bulge's K-band magnitudes.

\begin{acknowledgements}
The authors are very grateful to the anonymous referee for his/her
instructive comments that significantly improved the content of the
paper. We thank Luis C. Ho, Chien Y. Peng, and Leslie K. Hunt for
thoughtful discussion and kind help. QGU would like to acknowledge
the financial support from the China Scholarship Council (CSC). This
work is supported by the National Natural Science Foundation of
China under grants 10103001 and 10221001. This research made use of
the NASA/IPAC Extragalactic Database (NED), which is operated by the
Jet Propulsion Laboratory, California Institute of Technology, under
contract with the National Aeronautics and Space Administration.
This publication makes use of data products from the Two Micron All
Sky Survey, which is a joint project of the University of
Massachusetts and the Infrared Processing and Analysis
Center/California Institute of Technology, funded by the National
Aeronautics and Space Administration and the National Science
Foundation.

\end{acknowledgements}

%-------------------------------------------------------------------

\end{document}